\documentclass[11pt,a4paper]{article}
\pdfoutput=1
\usepackage{jheppub}
\usepackage[english]{babel}
\usepackage{amssymb,amsmath}
\usepackage[dvipsnames,x11names]{xcolor}
\usepackage{graphicx}
\usepackage{mathtools}
\usepackage{slashed}
\usepackage{float}
\usepackage{siunitx}
\usepackage{comment}
\usepackage[labelfont=normalfont,font=it]{caption}
\usepackage[normalem]{ulem}
\usepackage{dcolumn}
\usepackage{textcomp}
\usepackage{xfrac}
\usepackage{multirow}

\makeatletter
\gdef\@fpheader{}
\makeatother

\DeclarePairedDelimiter\abs{\lvert}{\rvert}%

\begin{document}

\preprint{%
  \parbox[t]{4.5cm}{\raggedleft
    IFT-UAM/CSIC-25-150\\
    MIT-CTP/5899\\
    KCL-PH-43}}

\title{A new connection between WIMP dark matter and the hierarchy problem}

\author[a]{Maximilian Detering,}
\author[b,c,d]{Thomas Steingasser}
\author[a]{and Tevong You}

\affiliation[a]{%
Theoretical Particle Physics and Cosmology Group, Department of Physics, \\ King’s College London, London, WC2R 2LS, UK
}%

\affiliation[b]{%
Departamento de Fisica Teorica, Universidad Autonoma de Madrid, and IFT-UAM/CSIC, Cantoblanco, 28049, Madrid, Spain
}%

\affiliation[c]{%
MIT Center for Theoretical Physics -- a Leinweber Institute, Massachusetts Institute of Technology, Cambridge, MA  02139, USA}

\affiliation[d]{Black Hole Initiative at Harvard University, 20 Garden Street, Cambridge, MA 02138, USA
}%

\emailAdd{maximilian.detering@kcl.ac.uk, thomas.steingasser@uam.es, tevong.you@kcl.ac.uk}

\date{\today}

\abstract{
This work proposes a direct link between the hierarchy problem and Weakly Interacting Massive Particles (WIMPs): we suggest that the small mass of the Higgs boson arises from being dynamically driven to the scale of the WIMP. Such a special electroweak vacuum is singled out by lying close to the critical boundary of a phase transition, as recently explored in a new class of cosmological solutions to the hierarchy problem. They generically predict the Higgs potential to be destabilised just above the weak scale. Intriguingly, the requirement for new physics to achieve this coincides with two independently well-motivated expectations: a split spectrum of light fermions and heavy bosons, as anticipated from naturalness, and the so-called ``WIMP miracle''. A WIMP with mass around the weak scale not only happens to have the correct thermal relic abundance to be the dark matter (DM), it can also give rise to the necessary critical boundary at the TeV scale through its Yukawa couplings to the Higgs. We use a higgsino-like singlet–doublet model to illustrate our Higgs-DM criticality scenario and show that if this WIMP DM mass is observed to be greater than $\sim 1.2$ TeV then it necessarily implies a strong bound on the Higgs mass and an upper bound on the scale of heavy new physics that restores vacuum stability. It can be thoroughly probed in direct detection experiments, astrophysical signals and future collider searches, further motivating a comprehensive exploration of the remaining heavy WIMP parameter space.}

\maketitle

\section{Introduction} 

Uncovering the nature of Dark Matter (DM) and understanding the hierarchy problem of the Higgs are two of the biggest mysteries in fundamental physics. Although there is overwhelming evidence for the existence of DM from multiple independent astrophysical and cosmological observations (see e.g.~Ref.~\cite{Cirelli:2024ssz} for a review), the underlying particle responsible has yet to be detected directly by experiments and its detailed properties are still unknown; notoriously, its mass could lie anywhere in the $\sim\!10^{-21}$ eV~\cite{Zimmermann:2024xvd} to super-Planckian range and the only certainty about its couplings is that it must interact gravitationally. Theoretical priors are therefore useful in identifying promising DM candidates to target.   

A compelling possibility singled out early on is the thermal Weakly Interacting Massive Particle (WIMP) paradigm~\cite{Feng:2022rxt}. The thermal freeze-out production mechanism has been spectacularly successful in explaining the cosmological history and abundance of our visible sector. It could furthermore account for the dark sector's relic abundance if the DM is a particle with mass around the weak scale and some degree of coupling to the Standard Model (SM), for example through the weak interactions. The fact that the thermal WIMP scenario points coincidentally towards new physics at the weak scale, where a resolution to the Higgs naturalness puzzle is also expected to emerge, has been dubbed the ``WIMP miracle". It is therefore not surprising to find that many WIMP DM candidates are naturally embedded in symmetry-based solutions to the hierarchy problem, such as supersymmetric or composite Higgs models. 

However, the lack of weak-scale new physics beyond the Standard Model (BSM) at the CERN Large Hadron Collider (LHC) and the absence of WIMPs in direct detection experiments have driven a healthy diversification of efforts into investigating alternative ideas for the origins of DM and the Higgs. It should nevertheless be emphasised that there is a crucial difference in null results in the search for WIMP DM and for new physics associated to the Higgs: naturalness places a fine-tuning prior favouring lighter masses on the parameter space of new physics expected to resolve the hierarchy problem, whereas WIMP DM could lie more or less anywhere within the range of masses compatible with being a thermal relic. WIMP DM therefore remains amongst the most attractive and motivated DM candidates despite null results in searches so far. On the other hand, the absence of the expected new physics in conventional solutions to the hierarchy problem calls for new ideas and alternative frameworks for understanding the mysterious Higgs sector~\cite{Giudice:2013yca, Giudice:2017pzm, Craig:2022eqo}.   

In this context, various cosmological mechanisms have been proposed to dynamically fix the Higgs mass to its observed value in the absence of new symmetries protecting it at the weak scale (see e.g.~Ref.~\cite{Asadi:2022njl} for a review). One such approach takes inspiration from Self-Organised Criticality (SOC)~\cite{Bak:1987xua}, an emergent phenomenon observed in many systems where the interplay of collective dynamics tunes itself towards a critical point separating two phases. Given the ubiquity of SOC and nature's tendency to repeat itself, it is not far-fetched to suppose that some form of SOC could also have taken place on cosmological scales~\cite{Giudice:2008bi}.

Several concrete mechanisms have been put forward. For example, stochastic dynamics during inflation can sample a wide variety of vacua with different properties. This provides the necessary ingredients for SOC to preferentially select an electroweak (EW) vacuum near criticality as the most likely universe to end up in after reheating. In \textit{Self-Organised Localisation} (SOL)~\cite{Giudice:2021viw}, fluctuations of a spectator scalar field scan over a parameter range with a critical point triggering a quantum phase transition, and under the right conditions the equilibrium scalar field distribution localises itself near that critical point. Another SOC mechanism studies more global dynamics of vacuum transitions in a landscape, finding that metastable vacua are the most likely to be selected~\cite{Khoury:2019yoo,Khoury:2019ajl,Kartvelishvili:2020thd,Khoury:2021grg}. These, in turn, have been shown to require an unnaturally small Higgs mass~\cite{Khoury:2021zao,Benevedes:2024tdq}. Many of these proposals require eternal inflation. As such, they are currently limited by our incomplete understanding of the inherent measure problem. Other possibilities bypass this requirement, relying on vacuum selection criteria other than SOC. For example, the recently proposed \textit{Higgs-Driven Crunching Mechanism} avoids the notion of probability entirely by introducing the near-criticality as a necessary condition for any vacuum to avoid cosmological crunching~\cite{Benevedes:2025qwt}.

Regardless of the specific underlying model of vacuum selection, which we remain agnostic about in this work, if the Higgs mass arises from living on the edge of a critical point at the TeV scale, then a generic prediction is new physics coupled to the Higgs that destabilises the Higgs potential. The vacuum instability scale must be in the TeV range such that a larger value of the Higgs mass would prevent the existence of the electroweak vacuum. If such a cosmological self-tuning of the Higgs mass is indeed part of the solution to the hierarchy problem, our vacuum will be observed to be metastable with a drastically different Higgs potential that should become apparent at TeV energy scales. This unavoidably implies accessible new physics coupling to the Higgs, but, crucially, of a different kind of structure to symmetry-based models addressing the hierarchy problem.       

Remarkably, the form of new physics required for a metastable Higgs with a lowered vacuum instability scale is precisely that of a split spectrum expected from naturalness arguments: vector-like (VL) fermions, which destabilise the Higgs through their Yukawa couplings, can be naturally light, while unprotected scalars (other than the Higgs) and gauge bosons, which restore the stability of the Higgs potential, are expected to be naturally heavy. Moreover, we have independent motivation for the possible existence of vector-like fermions around the EW scale: the thermal WIMP DM paradigm. In our Higgs-DM criticality mechanism, the WIMP plays a central role in setting the metastability bound on the Higgs mass and hence its value at the EW scale, as cosmological dynamics in the early universe drive the Higgs mass to be in proximity to the WIMP. The WIMP miracle may thus be more than a mere coincidence of scales --- we suggest that it is directly responsible for the hierarchy between the observed Higgs mass and the heavy scale of new physics that this mass would naturally have been at.

\begin{figure}[t!]
\centering
\includegraphics[width = 1.0\textwidth]{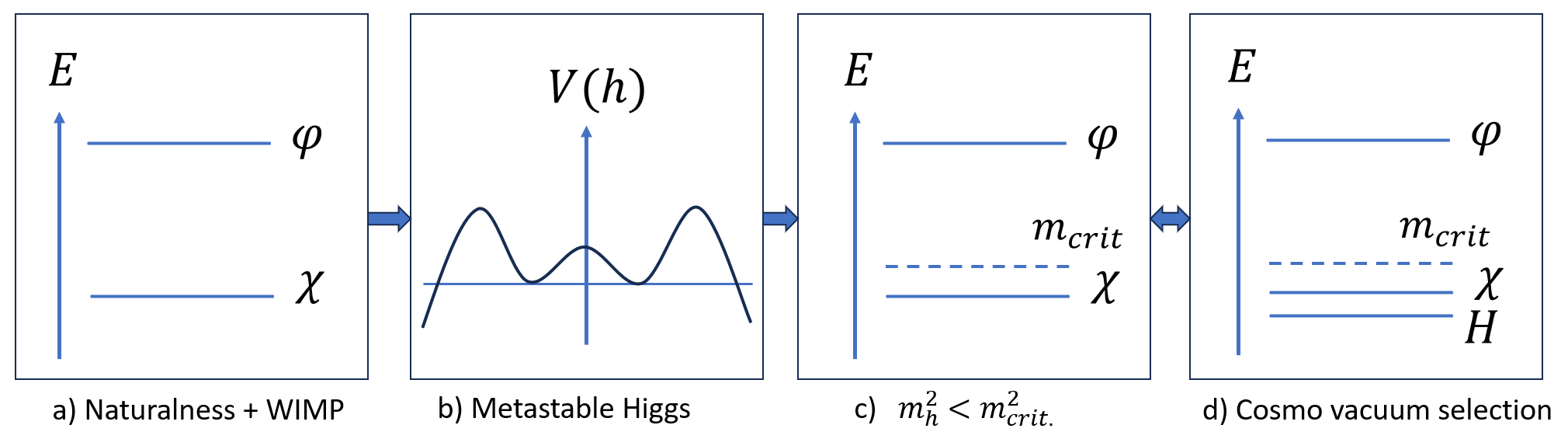}
\caption{Schematic summary of our near-critical Higgs DM scenario in which the Higgs mass is dynamically driven to the WIMP Scale. a) A split spectrum of naturally light fermions and heavy bosons is expected from naturalness, and thermal WIMP DM further motivates the existence of fermions around the weak scale. b) Such a split spectrum can destabilise the Higgs potential, since fermion Yukawa couplings contribute negatively to the renormalisation group running of the Higgs quartic while positive contributions from bosons are decoupled to heavier scales. c) This leads to an upper bound on the Higgs mass in order for our electroweak vacuum to exist. d) An underlying cosmological mechanism places the Higgs mass close to this critical boundary.}
\label{fig:mechanism_summary}
\end{figure}

A schematic of our critical Higgs DM scenario is summarised in Fig.~\ref{fig:mechanism_summary}. This general picture may be realised in many different ways. We choose here a singlet-doublet model of WIMP DM as a simplified example. Such a singlet-doublet WIMP corresponds to a higgsino in a supersymmetric framework, though in our case we will see that a Yukawa coupling larger than that of the higgsino is required for our mechanism. The larger Yukawa coupling also places this simplified model under more tension from direct detection experiments. Nevertheless, this simplified model serves to illustrate qualitatively the relations between the WIMP DM mass, the vacuum instability scale, and the heavy scale of new physics that restores vacuum stability and we show that there remains viable parameter space that can be comprehensively explored by future searches.

This article is organised as follows. In Sec.~\ref{sec:model} we introduce the singlet-doublet WIMP model and set our notation. In Sec.~\ref{sec:DMMSB} we discuss the relation between WIMP DM and Higgs criticality and identify the overlap in the parameter space for the thermal relic abundance and the vacuum metastability bound on the Higgs mass. Sec.~\ref{sec:lifetime} studies the lifetime of this metastable electroweak vacuum and computes the upper bound on the scale of new physics necessary to restore vacuum stability sufficiently for compatibility with the observed lifetime of our universe. Finally, Sec.~\ref{sec:constraints} considers astrophysical, collider, and direct detection constraints on the singlet-doublet DM model. We conclude in Sec.~\ref{sec:conclusion} with some general remarks.

\section{The singlet-doublet WIMP model}\label{sec:model}

The singlet-doublet model (SDM) is a minimal vector-like extension of the SM with new EW states that include a WIMP DM candidate. It can be found in more complete ultraviolet (UV) theories such as, for example, the bino-higgsino of supersymmetric models where the Yukawa couplings are fixed by supersymmetry to be related to the gauge couplings. We make no such assumptions here and treat the SDM as a simplified toy model. See, e.g.\@, Refs.~\cite{Mahbubani:2005pt, Cohen:2011ec, Cheung:2012qy, Cheung:2013dua, Yaguna:2015mva, Han:2018gej, Arcadi:2019lka, Paul:2024prs, Bhattiprolu:2025beq, Arcadi:2025sxc} for dedicated studies of this WIMP DM model.

The model includes a singlet $\psi$ transforming as a left-handed Weyl spinor in the Standard Model ($\mathrm{SU}(3)_{c}\times\mathrm{SU}(2)_{L}\times\mathrm{U}(1)_{Y}$) gauge representation $\left(\mathbf{1},\mathbf{1},0\right)$. Furthermore, it contains a vector-like pair of left-handed $\mathrm{SU}(2)_{L}$ doublets, $\chi_d, \chi_u$, transforming in the representations $\left(\mathbf{1},\mathbf{2},-\frac{1}{2}\right)$ and $\left(\mathbf{1},\mathbf{2},+\frac{1}{2}\right)$ of the SM respectively. In particular, we decompose the doublets into two left-handed Weyl spinors as\footnote{Note the additional minus sign in the lower component of the negative hypercharge doublet, which has been written explicitly to obtain the canonical sign for Dirac mass term later in Eq.~\eqref{eq:mass-lagrangian}. We can choose this sign by a chiral rotation since only the combination of the phase of the mass and the spinor is physical.}
\begin{equation}
    \chi_d = \begin{pmatrix}\chi_{d}^{0} \\ - \chi_{d}^{-}\end{pmatrix}
    ,\quad
    \chi_u = \begin{pmatrix}\chi_{u}^{+} \\ \chi_{u}^{0}\end{pmatrix}
    .
    \label{eq:doublets}
\end{equation}
The singlet-doublet model is described by the Lagrangian,
\begin{equation}\label{eq:lagrangian-sdmodel}
\begin{split}
    \mathcal{L}_\mathrm{SD} = \;
    & i\Bar{\psi} \slashed{\partial} \psi
    + i\Bar{\chi}_u \slashed{D} \chi_u
    + i\Bar{\chi}_d \slashed{D} \chi_d
    \\
    & - \frac{1}{2} m_S \psi^{c} \psi
    - m_D \chi_d \cdot \chi_u
    - y_1 \chi_u H^{\dagger} \psi
    - y_2 \chi_d \cdot H \psi
    + \text{h.c.} ,
\end{split}
\end{equation}
where the dot product implies the $\mathrm{SU}(2)$ invariant contraction via the antisymmetric tensor and $H$ is the Higgs doublet.
This is the most general extension that respects a $\mathbb{Z}_2$ symmetry under which the SM particles are even and the singlet and doublets are odd. To parametrise the Yukawa structure, we define $y_1 = y \, \cos{\theta}$ and $y_2 = y \, \sin{\theta}$. The singlet mixes with the two neutral components of the doublets into three Majorana mass eigenstates. The two electrically charged Weyl spinors of the doublets, transforming in opposite representations of hypercharge, combine into a Dirac spinor. The mass terms of the singlet-doublet model in terms of Weyl spinors read after electroweak symmetry breaking (where in unitary gauge $H = 1/\sqrt{2} (0,v+h)^T$),
\begin{equation}
\begin{split}
    \mathcal{L}_\text{mass} = -\frac{1}{2} m_S \psi^{c} \psi - m_D \chi_{u}^{+} \chi_{d}^{-} - m_D \chi_{u}^{0} \chi_{d}^{0} - \frac{y_1 v}{\sqrt{2}} \chi_{u}^{0} \psi - \frac{y_2 v}{\sqrt{2}} \chi_{d}^{0} \psi + \text{h.c.} \, ,
\end{split}
\label{eq:mass-lagrangian}
\end{equation}
and we embed the Weyl spinors into three Majorana spinors $\Psi_i$ and a Dirac spinor $X$ as follows,
\begin{equation}\label{eq:Majorana-embedding}
    \Psi_1 = \begin{pmatrix} \psi \\ -i\sigma_2 \psi^* \end{pmatrix} , \;
    \Psi_2 = \begin{pmatrix} \chi_u^0 \\ -i\sigma_2 {(\chi_u^0)}^* \end{pmatrix} , \;
    \Psi_3 = \begin{pmatrix} \chi_d^0 \\ -i\sigma_2 {(\chi_d^0)}^* \end{pmatrix} , \;
    X = \begin{pmatrix}
        \chi_u^+ \\ -i\sigma_2 {(\chi_d^-)}^*
    \end{pmatrix} .
\end{equation}

\paragraph{Masses and mixing}
The Dirac fermion mass is given by $m_D$ at tree-level.
From Eq.~\eqref{eq:mass-lagrangian}, we can read off the mass matrix for the three Majorana spinors (or here in terms of the associated left-handed Weyl spinors $\psi_i$) following Eq.~\eqref{eq:Majorana-embedding},
\begin{equation}\label{eq:massmatrix}
    \mathcal{L}_{\text{mass},\psi} = - \frac{1}{2} \boldsymbol{\psi}^T \mathcal{M} \boldsymbol{\psi} + \text{h.c.}
    \quad \text{with} \quad
    \mathcal{M} =
    \begin{pmatrix}
        m_S & \frac{y_1 v}{\sqrt{2}} & \frac{y_2 v}{\sqrt{2}} \\
        \frac{y_1 v}{\sqrt{2}} & 0 & m_D \\
        \frac{y_2 v}{\sqrt{2}} & m_D & 0
    \end{pmatrix}
    .
\end{equation}
We can then find the mass basis of the Majorana spinors, $\mathbf{N} = (N_1, N_2, N_3)^T$. The mass matrix $\mathcal{M}$ can be unitarily diagonalised to a generally indefinite diagonal matrix. However, the Majorana mass terms are only defined up to a chiral rotation. We are therefore looking for the Takagi factorisation of the complex symmetric mass matrix $\mathcal{M}$ such that $D$ is a real positive semi-definite diagonal matrix,
\begin{equation}
    \mathrm{diag}(m_1, m_2, m_3) \equiv D = U^T \mathcal{M} U
    \quad \text{for} \quad \mathbf{\Psi} = \mathcal{U} \mathbf{N} \quad \text{with} \quad \mathcal{U} = \begin{pmatrix}
        U & 0 \\
        0 & U^{*}
    \end{pmatrix}.
\end{equation}
Here, $\mathcal{U}$ is acting on the left- and right-handed Weyl spinors, and $U$ is a unitary matrix. Without loss of generality, we assume $m_1 < m_2 < m_3$ such that $m_{\rm DM} = m_1$ is the mass of the DM candidate. For simplicity, we also assume all parameters in the Lagrangian as written in Eq.~\eqref{eq:lagrangian-sdmodel} to be real, though in general, not all complex phases can be absorbed. The mass eigenstates are therefore the admixture of the singlet and doublet as (in the left-handed Weyl spinor basis)
\begin{equation}
    N_{Li} = (U^{\dagger})_{ij} \Psi_{Lj}.
\end{equation}
The amount of mixing between the singlet and doublets plays a crucial role for the phenomenology. The singlet fraction in the DM candidate is given by $\abs{U_{11}}^2$, while the doublet fraction is given by $\abs{U_{21}}^2 + \abs{U_{31}}^2$, and unitarity of $U$ ensures $\abs{U_{11}}^2 + \abs{U_{21}}^2 + \abs{U_{31}}^2 = 1$. For heavy singlet and doublet masses above the EW scale, the amount of mixing is mainly determined by the ratio $m_S/m_D$. The case $m_S/m_D \gg 1$ corresponds to mostly-doublet DM, and  $m_S/m_D \ll 1$ yields mostly-singlet DM. For masses $m_S \approx m_D$ above the EW scale, the DM is an equal mixture of singlet and doublet components.

\paragraph{Interactions}
The dark sector interacts with the SM through all bosons. In the mass basis for the Majorana and Dirac fermions, the interactions are given by the following Lagrangian, 
\begin{align}
    \mathcal{L}_\mathrm{int} = & 
    + e \Bar{X} \slashed{A} X
    + g_L \Bar{X} \slashed{Z} X + \Bar{X} W^{+}_\mu \left(g_{V}^{X N_i W} \gamma^\mu + g_{A}^{X N_i W} \gamma^\mu \gamma^5 \right) N_i \nonumber\\
    &- g_{A}^{N_i N_j Z} \Bar{N}_i \slashed{Z} \gamma^5 N_j -  y_{h\chi_i \chi_j} h \Bar{N}_i N_j
    + \text{h.c.} \; .
\end{align}
We define the above couplings as
\begin{align}
    g_A &= \frac{\sqrt{g^2 + g'^2}}{4}, &
    g_L &= \frac{1}{2} \frac{g^2 - {g^\prime}^2}{\sqrt{g^2 + {g^\prime}^2}}, \\
    g_{V}^{X N_i W} &= \frac{g}{2\sqrt{2}}( U_{i3} + U_{i2} ), & 
    g_{A}^{X N_i W} &= \frac{g}{2\sqrt{2}} (U_{i3} - U_{i2}), \\
    g_{A}^{N_i N_j Z} &= g_A \left( U^{*}_{i3} U_{j3} - U^{*}_{i2} U_{j2} \right), & 
    y_{h\chi_i \chi_j} &= \frac{y_1}{\sqrt{2}} U^{*}_{i2} U_{j1} + \frac{y_2}{\sqrt{2}} U^{*}_{i3} U_{j1},
\end{align}
where the couplings are not yet symmetrised in the Majorana indices.
It is straightforward to show that the coupling of the DM candidate to the $Z$ boson, $g_{A}^{N_1 N_1 Z}$, vanishes in the $Z$ blind spot characterised  by $\tan{\theta} = -1$, and similarly its coupling to the Higgs boson vanishes in the Higgs blind spot in the mostly-doublet regime, characterised by $\tan{\theta}=-1$ and $m_S > m_D$. These blind spot regions have been studied most recently in Refs.~\cite{Han:2018gej, Bhattiprolu:2025beq, Arcadi:2025sxc}. We will focus on this region of parameter space to evade direct detection constraints, though we note that this tuning may be independently justified or related to other aspects of the model in more UV-complete theories. For example, in models of split supersymmetry or the NMSSM this choice follows from the requirement of a small Higgs mass~\cite{Cheung:2012qy}~\footnote{We note that any potential embedding of our mechanism in a supersymmetric framework would be at the heavy cut-off scale such that the hierarchy between this heavy scale and the electroweak scale is not protected by the usual symmetry mechanisms. }.

\section{Dark matter and Higgs criticality}\label{sec:DMMSB}

The singlet-doublet model has the characteristic features to accommodate both a WIMP DM candidate and a Yukawa coupling to the Higgs that is necessary to lower the instability scale of the EW vacuum by modifying the renormalisation group (RG) evolution of the Higgs self-coupling compared to the SM. As we will discuss below, this may have important implications for the hierarchy problem. Those two aspects of the singlet-doublet model---dark matter and Higgs criticality---target orthogonal directions in the model parameter space; thus, taken together, they single out a particular region of the parameter space. In the following, we will first review the DM relic density in the singlet-doublet model before discussing how it realises Higgs criticality.

\subsection{Dark matter relic abundance}\label{sec:dm-abundance}

The WIMP DM candidate in the singlet-doublet model is a Majorana fermion that couples through gauge and Higgs interactions to the SM. These couplings ensure that the BSM sector was initially in thermal equilibrium with the SM. Hence, the DM relic abundance is set through the standard freeze-out mechanism.

We compute the DM relic abundance using \textsc{micrOMEGAs v6.2.4} \cite{Alguero:2023zol} with a custom model file written with \textsc{FeynRules v2.3} \cite{Alloul:2013bka} used to generate \textsc{CalcHEP} \cite{Belyaev:2012qa} model files. In \textsc{micrOMEGAs}, we identify the DM candidate and numerically solve the Boltzmann equation for freeze-out\footnote{We use the fast calculation option and work in Feynman gauge. We neglect contributions from off-shell gauge bosons in the final state.}. For the scan over the parameter space, we numerically diagonalise the mass matrix at each model parameter point.
The DM relic abundance predicted in the SDM sensitively depends on the model parameters, most notably on the singlet and doublet masses, $m_S$ and $m_D$. The observed DM density is measured to be \cite{Planck:2018vyg},
\begin{equation}
    \Omega_{\mathrm{DM}}h^2 = \num{0.120(1)}.
\end{equation}
The SDM as a DM model would therefore be excluded in the part of the parameter space that leads to an overproduction of DM. 
Furthermore, as we will discuss in Sec.~\ref{sec:constraints}, constraints from direct collider searches and precision tests as well as direct and indirect detection experiments severely constrain the parameter space. In particular, we consider the scenario of mostly-doublet DM with $m_S > m_D$ in the following. We will focus for simplicity on the parameter space in the $Z$ and Higgs boson blind spot region with $y_1=-y_2$, i.e.\@ $\tan{\theta}=-1$ that evades direct detection constraints; see e.g.~Ref.~\cite{Bhattiprolu:2025beq} for a more general analysis of singlet-doublet DM.

We can understand the parametric dependence of the DM relic abundance in the following way. The relic abundance is set by the annihilation cross section of the DM fermions into SM particles, and for a thermal target a velocity-averaged cross section of~\cite{Cirelli:2024ssz}
\begin{equation}
    \left<\sigma v \right> \sim \SI{e-26}{\cubic\centi\meter\per\second}
    \label{eq:relicxsection}
\end{equation}
yields the observed DM abundance. In the blind spot region we are considering here, the diagonal couplings of the DM candidate to the Higgs boson and $Z$ boson vanish. Thus, the dominant contribution to the annihilation cross section stems from $t$- and $u$-channel exchange of $N_2, N_3$ or $X$ with $ZZ$ and $WW$ final states respectively. The annihilation strength is therefore set by the gauge interactions and largely unaffected by the size of the Yukawa coupling for masses above the EW scale. Instead, the annihilation cross section is mainly sensitive to the DM mass, which in the mostly-doublet regime, $m_S > m_D$, is just controlled by the doublet mass. 

\begin{figure}[t]
    \centering
    \includegraphics{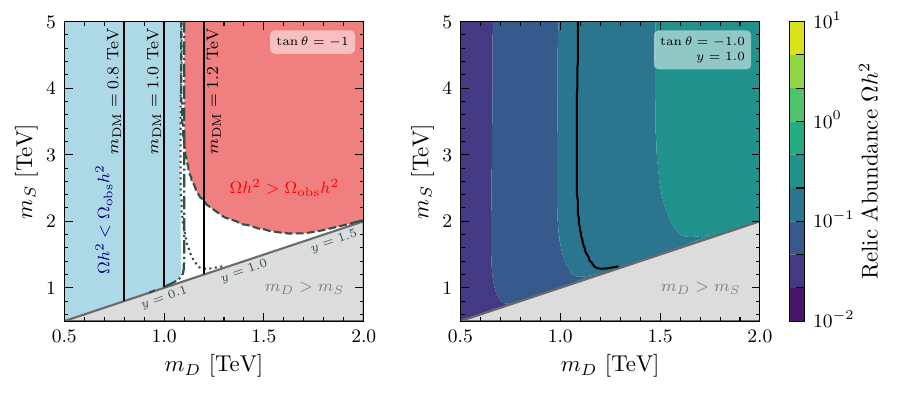}
    \caption{Thermal dark matter relic density $\Omega_\text{DM}h^2$ in the blind spot with $\tan{\theta}=-1$. \textbf{Left:} Regions of the parameter space leading to DM over-production for $y=1.5$ (red) and under-production for $y=0.1$ (blue). The contours corresponding to the observed DM abundance are shown for Yukawa couplings $y=0.1$ (dash-dotted), $y=1.0$ (dotted) and $y=1.5$ (dashed). Contours of DM mass, $m_\mathrm{DM}$, in the mostly-doublet regime are represented by solid black lines. \textbf{Right:} Relic abundance contours for $y=1.0$ in the mostly-doublet regime. The black solid line corresponds to the observed relic density.}
    \label{fig:relic-density}
\end{figure}

We show the DM relic abundance $\Omega_{\mathrm{DM}} h^2$ for a Yukawa coupling $y=1.0$ in the blind spot region of $\tan{\theta}=-1$ in the right panel of Fig.~\ref{fig:relic-density}. The observed relic density is indicated by a solid black line. In the left panel of Fig.~\ref{fig:relic-density}, we show in solid black the DM mass $m_{\rm DM}$ as a function of the singlet and doublet masses for $\tan{\theta}=-1$. 
We consider the DM relic abundance for Yukawa couplings between 0.1 and 1.5 and find that too low a doublet mass $m_D$ corresponds to an insufficient production of DM, as denoted by the blue region for $y=0.1$, while large doublet masses lead to an overproduction of DM, as shown by the red region for $y=1.5$. We see that for $m_S \gg m_D$, a doublet mass of $m_D \simeq \SI{1.1}{TeV}$ yields the observed DM abundance largely independent of the Yukawa coupling \cite{Abe:2025lci}. However, if $m_S \sim m_D$, the DM relic abundance is sensitive to the Yukawa coupling. The part of the parameter space that can give the observed DM abundance for $0.1 \leq y \leq 1.5$ corresponds to the white region in the left panel of Fig.~\ref{fig:relic-density}. The DM relic density contours for the values $y=0.1$, $1.0$ and $1.5$ are shown in solid, dotted, and dashed grey lines, respectively. The greyed out region is where $m_D > m_S$ that is mostly excluded by direct detection, as we discuss in Sec.~\ref{sec:constraints}.

\subsection{Higgs mass metastability bound}\label{sec:msbound}

\begin{figure}[t!]
    \centering
    \includegraphics[width = 0.48\textwidth]{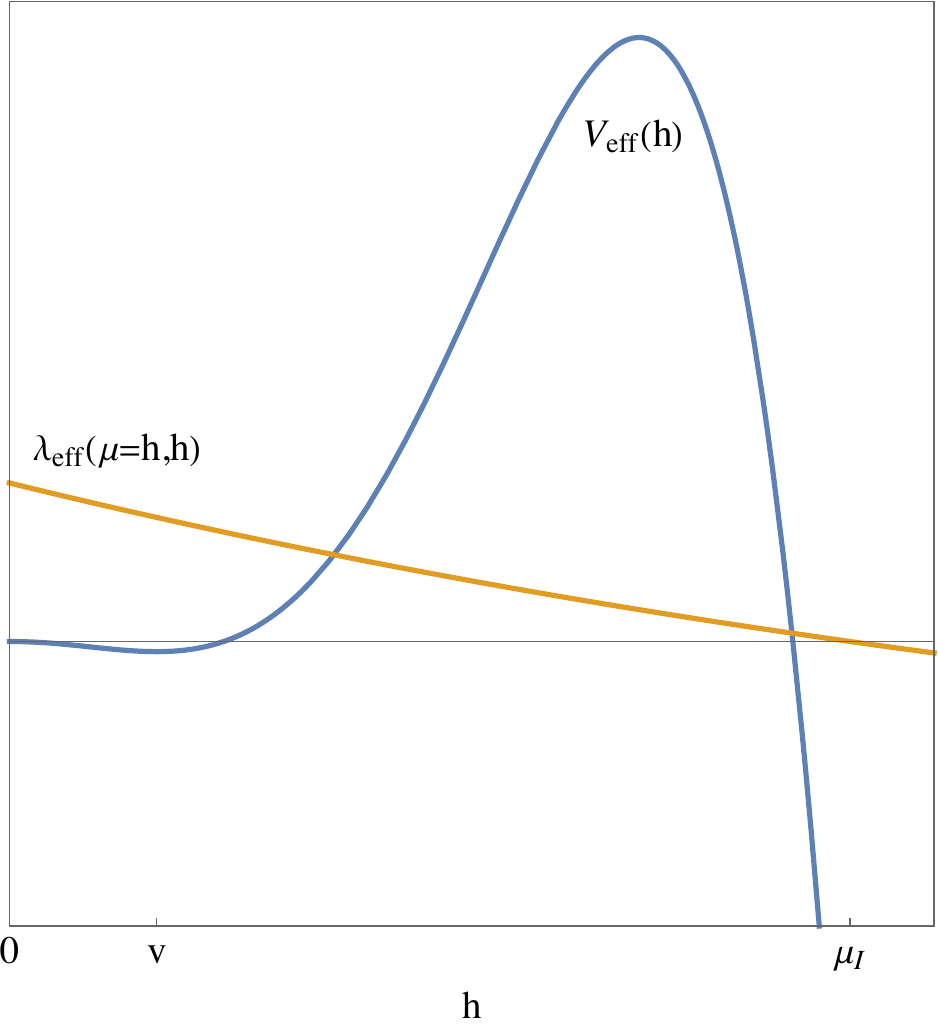}
    \includegraphics[width = 0.48\textwidth]{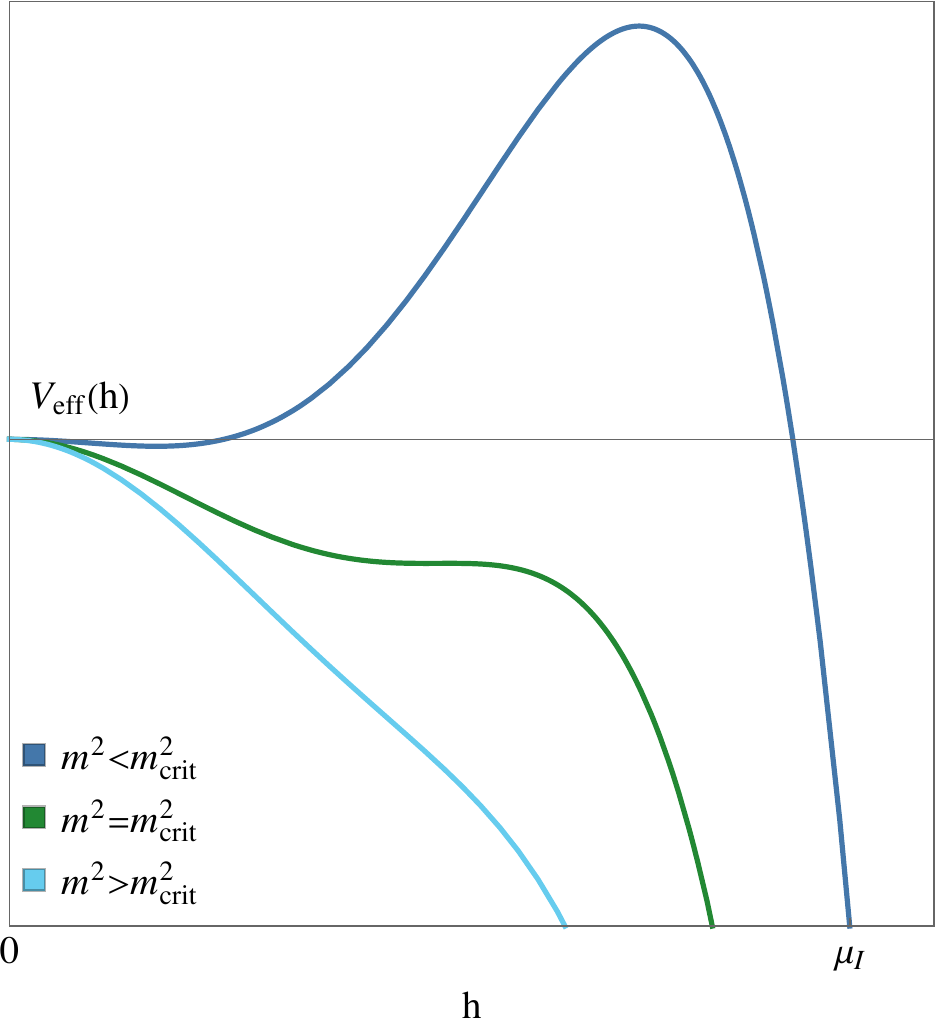}
    \caption{\textbf{Left:} Sketch of the effective Higgs potential $V_{\rm eff}$ in Eq.~\eqref{eq:Veff} together with the effective quartic coupling $\lambda_{\rm eff}$. The coupling crossing zero due to its RG evolution leads to the potential turning over. \textbf{Right:} Sketch of the effective potential for a fixed RG-trajectory of $\lambda$ and three different values of the mass parameter.}
    \label{fig:scales}
\end{figure}

An essential feature of the WIMPs discussed in Sec.~\ref{sec:dm-abundance} are their Yukawa couplings to the Higgs, contributing to their mass and mixing. Adding to their phenomenological relevance, they also influence the shape of the Higgs potential at high energies. This can be traced back to loop corrections to the Higgs quartic self-coupling $\lambda$ whose running is controlled by its beta function $\beta_\lambda = {\rm d} \lambda/ {\rm d} \ln \mu$. Introducing the BSM fermions manifests in a one-loop contribution to this function of the form 
\begin{align}
    \Delta \beta_\lambda^{(1)} = \frac{1}{(4 \pi)^2}  \left[ -2 (y_1^2 + y_2^2)^2 + \lambda (4 y_1^2 + 4 y_2^2)\right] .
\end{align}
For large enough values of the Yukawa couplings, this contribution is dominated by the first \emph{negative} term, driving $\lambda$ to smaller values as the RG scale is increased. This pushes $\lambda$ to vanish and consequently take a negative value above the so-called \textit{instability scale}. Throughout this article, we define the instability scale through the full loop-corrected quartic coupling, $\lambda_{\rm eff}(\mu=\mu_I, h=\mu_I)=0$. In the SM the instability scale is at $\sim 10^{10}$ GeV, too high for our purpose. However, for large enough Yukawa couplings, this scale can be as low as $\mathcal{O}(1-100)$ TeV. This effect also influences the effective potential, which is of the form
\begin{align}
    V_{\rm eff}(h)= - \frac{1}{4}m_{\rm eff}^2(\mu=h,h)h^2+ \frac{1}{4} \lambda_{\rm eff}(\mu=h,h)h^4.
    \label{eq:Veff}
\end{align}
Due to the sign of the mass term, $\lambda$ changing its sign amounts to the potential flipping over. The behaviour of the effective potential as a function of the effective quartic coupling and mass parameters is illustrated in Fig.~\ref{fig:scales}. 

However, this shape of the potential relies on the implicit assumption that there exists some regime in which the potential is dominated by the quartic term and for which $\lambda>0$. The requirement for the quartic term to dominate is characterised by field values of order $\phi^2 \gtrsim m^2/\lambda$, while $\lambda$ is only positive for $\phi \lesssim \mu_I$. From these two conditions, we see that this regime can only exist for mass parameters smaller or at best comparable to the instability scale. This is illustrated in Fig.~\ref{fig:scales}: if the mass parameter is too large, the quartic term only becomes dominant at field values for which $\lambda$ is negative, preventing the formation of the EW vacuum. More precisely, the \textit{Higgs mass metastability bound} can be calculated to be~\cite{Buttazzo:2013uya,Khoury:2021zao} 
\begin{align}\label{eq:msbound}
    m_h^2< m_{\rm crit}^2 \equiv e^{-3/2} |\beta_\lambda (\mu_I)| \mu_I^2.
\end{align}
For a detailed derivation of this inequality, see Ref.~\cite{Khoury:2021zao}. Note that we have neglected the running of the mass parameter as it is subdominant, being renormalised multiplicatively with $\beta_{m_h^2}\propto m_h^2$.

A series of recent works explored the possibility to invoke this bound as a solution to the long-standing hierarchy problem by proposing a justification for the Higgs mass to be near this critical value. First, Ref.~\cite{Giudice:2021viw} considered a simple model in which the Higgs mass is scanned by an additional scalar field (that is protected by a shift symmetry and so is technically naturally light). Given a suitable RG trajectory of the quartic coupling and potential, the authors show that fluctuations during eternal inflation naturally drive the scalar to a value close to saturating the bound of Eq.~\eqref{eq:msbound}. Meanwhile, Ref.~\cite{Khoury:2021zao} argues that Eq.~\eqref{eq:msbound} is a necessary condition for any metastable vacuum capable of reproducing the symmetry breaking pattern of the SM, i.e., with a negative quadratic term.\footnote{A similar argument can also be used to constrain the value of the mass parameter in the case of a positive quadratic term~\cite{Benevedes:2024tdq}.} This shifts the task of explaining the hierarchy problem to that of explaining metastability of the EW vacuum, which can, e.g.\@, be achieved through the arguments presented in Refs.~\cite{Khoury:2019yoo,Khoury:2019ajl,Kartvelishvili:2020thd,Khoury:2021grg} or, alternatively, Ref.~\cite{Benevedes:2025qwt}. An important feature of this perspective is that it naturally relates the instability scale $\mu_I$ to the scale of new physics, $\Lambda$, yielding the natural expectation for the mass parameter. We revisit this connection in more detail in Sec.~\ref{sec:lifetime}. For a review of these works see Refs.~\cite{Steingasser:2023ugv,Steingasser:2024hqi, Detering:2025nmh}.

\begin{figure}[t]
    \centering
    \includegraphics{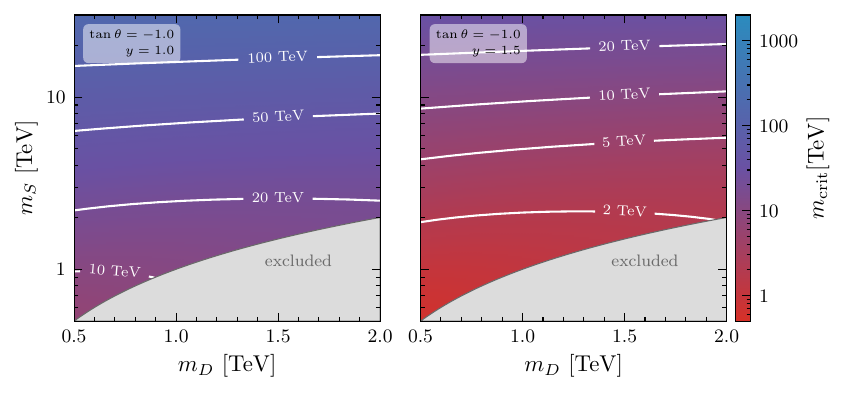}
    \caption{Metastability bound $m_\mathrm{crit}$ in the singlet and doublet mass plane for Yukawa couplings of $y=1$ (left) and $y=1.5$ (right) for $\tan{\theta}=-1$. The grey shaded region is mainly excluded by direct detection.}
    \label{fig:mD_vs_mS}
\end{figure}%

For the purpose of the present analysis, we remain agnostic about the underlying cosmological mechanism that selects a near-critical EW vacuum with a Higgs mass close to the bound and focus exclusively on how this metastability bound may be realised. Its qualitative behaviour can be understood in a straightforward manner through Eq.~\eqref{eq:msbound}: larger Yukawa couplings lead to a faster running of $\lambda$ towards negative values, thus lowering the instability scale $\mu_I$. Due to the form of the Yukawa interaction (see Eq.~\eqref{eq:lagrangian-sdmodel}), this effect is only present if both the singlet and doublet are not decoupled, i.e.~for scales $\mu$ where $\mu^2\gtrsim m_S^2, m_D^2$.\footnote{In the regime $m_D\lesssim \mu \ll m_S$ a similar effect takes place through effective interactions obtained by integrating out $m_S$ but these are suppressed by powers of $m_D/m_S$.} Thus, lighter fermions allow for a stronger metastability bound as low as $\mathcal{O}({\rm TeV})$ if the singlet and doublet masses are of the same order at that scale. The remaining little hierarchy between the metastability bound and the observed Higgs mass value can be regarded as an accident of coarse vacuum selection or explained by the particularities of this selection process that depend on the underlying cosmological mechanism. Moreover, we would not expect to be at the bound since there would then be no potential barrier at all. We refer the reader to the literature referenced above for more detailed discussion.

We note that the VL fermions with Yukawa couplings can get a sizeable mass contribution from the vacuum expectation value (VEV) of the Higgs if the latter takes on large values in the underlying cosmological vacuum selection mechanism. This, however, does not affect the applicability of the metastability bound to bounding the Higgs mass from above. The boundary condition of our effective field theory (EFT) description is for the quartic coupling to be negative at the natural heavy cut-off scale $\Lambda$, $\lambda(\Lambda) < 0$. If the VL fermion is too heavy, the running of $\lambda$ in the energy distance between $\Lambda$ and the mass of the VL fermion over which the VL fermion makes a large contribution is too small for it to have a significant effect. In this case $\lambda$ stays negative at energies below the VL fermion mass and the EW vacuum in the infrared (IR) will not exist. Only when the VL fermions are sufficiently light can their accumulated running contributions from $\Lambda$ to the VL fermion threshold be large enough to push the quartic coupling to positive values, giving rise to a metastable EW vacuum. For large VEV values above the EW scale the coexistence of the EW vacuum in the IR and the UV vacuum will therefore cease and leave only the UV vacuum, which is the necessary condition for the underlying cosmological vacuum selection to operate.

For example, in SOL the localisation near criticality depends only on the phases of the EW vacuum triggered by the scalar scanning the Higgs VEV. As argued above, the VL fermion with too heavy a mass from sampling large VEV values leads to the disappearance of the EW vacuum in the IR. Fig.~9 of Ref.~\cite{Giudice:2021viw} shows the resulting phase diagram of the scanning scalar in the case of a SOL mechanism localising the VEV at the boundary of the coexistence of the two phases. Similarly, in the case of Higgs-driven crunching the negative value of $\lambda$ at the UV matching scale arises naturally from the assumption that the EW vacuum emerges as a radiatively-generated vacuum ``on top'' of a local maximum.

\begin{figure}[t]
    \centering
    \includegraphics{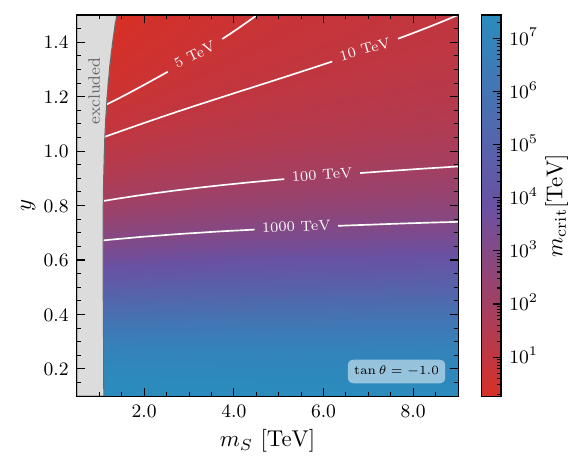}
    \caption{Critical value of the Higgs mass parameter $m_\mathrm{crit}$ as a function of the Yukawa coupling and singlet mass of the singlet-doublet model. We consider the mostly-doublet DM regime with $m_D < m_S$, and choose the correspondingly lightest doublet mass giving rise to the observed DM relic abundance. The grey region is excluded by direct detection constraints or cannot reproduce the observed DM abundance.}
    \label{fig:mD_vs_y}
\end{figure}

\subsection{WIMP dark matter and the instability scale}\label{sec:masses}

So far, we have found that the BSM fermions of the singlet-doublet model as described by Eq.~\eqref{eq:lagrangian-sdmodel} determine both the DM abundance and mass $m_{\rm DM}$ as well as as the critical Higgs mass $m_{\rm crit}$ setting the metastability bound. This suggests a correlation between these two masses, which we will explore in the following. We will find that the regime of a strong Higgs mass metastability bound corresponds to a distinctive range of DM masses, allowing for an indirect experimental verification of its possible importance through a DM measurement.

As laid out in Sec.~\ref{sec:dm-abundance}, we can identify the DM WIMP with the lightest mass eigenstate of the singlet-doublet system. In the blind spot region, the mass matrix~\eqref{eq:massmatrix} simplifies to
\begin{align}
\mathcal{M}_{\rm blind} =
    \begin{pmatrix}
        m_S & \frac{y v}{2} & -\frac{y v}{2} \\
        \frac{y v}{2} & 0 & m_D \\
        -\frac{y v}{2} & m_D & 0
    \end{pmatrix}
    .
\end{align}
To leading order in $v^2/m_S^2$, the absolute values of its eigenvalues are given by
\begin{align}
	m_1=m_D, \quad m_2\simeq m_D + \frac{v^2 y^2}{2(m_D + m_S)}, \quad m_3 \simeq m_S+ \frac{v^2 y^2}{2(m_D + m_S)} .
\end{align}
If $m_S>m_D$, as assumed throughout this work, this allows us to immediately identify $m_1$ as the smallest mass, and thus its corresponding state as the dark matter particle, $m_{\rm DM}=m_1$.

Following our discussion in Sec.~\ref{sec:msbound}, we can furthermore obtain a leading order estimate of the critical Higgs mass in terms of the model parameters. Following Eq.~\eqref{eq:msbound}, $m_{\rm crit}$ is dominated by the instability scale $\mu_I$, which is defined by the condition
\begin{align}\label{eq:minslin}
    0=\lambda_{\rm eff}(\mu_I,h=\mu_I)\simeq \lambda (\mu_I)+\delta \lambda_{\rm 1-loop}(\mu_I,h=\mu_I) .
\end{align}
Given a rapid running due to $\mathcal{O}(1)$ Yukawa couplings, we can linearise the RG evolution of $\lambda$ between the matching scale $m_S$ and the instability scale $\mu_I$. This allows us to approximate Eq.~\eqref{eq:minslin} as
\begin{align}
    0 &\simeq \lambda (m_S) -| \beta_\lambda (m_S) | \ln \left( \frac{\mu_I}{m_S} \right) +\delta \lambda_{\rm 1-loop}(\mu_I,h=\mu_I) \\ 
    \Rightarrow \quad \mu_I &\simeq m_S \cdot \exp \left( \frac{\lambda_{\rm eff}(m_S,h=m_S)}{|\beta_\lambda (m_S)|}\right) \label{eq:muinsLO}
\end{align}
Our discussion in Sec.~\ref{sec:dm-abundance} implies that the hypothesis that the lightest of the mass eigenstates accounts for the entire observed abundance of DM is only true on a one-dimensional contour in the $(m_D,m_S)$ parameter space. See Fig.~\ref{fig:relic-density}. 

From Fig.~\ref{fig:relic-density} and Eq.~\eqref{eq:muinsLO}, we see that a DM mass above a threshold of roughly $m_{\rm DM}\sim 1.2$~TeV can be considered a smoking gun for a strong metastability bound in this model. For $m_S \gg m_D\sim \mathcal{O}({\rm TeV})$, the DM mass is mostly independent of both $m_S$ and the Yukawa coupling $y$, varying only in a narrow window $m_{\rm DM}\simeq \SIrange[range-phrase = {-}, range-units = single]{1080}{1100}{GeV}$. Following from Eq.~\eqref{eq:muinsLO}, this is equivalent to a weak metastability bound. Conversely, a strong metastability bound is possible only in the regime $m_S\sim m_D \sim \mathcal{O}({\rm TeV})$. In this regime however, we find a significant dependence of $m_S$ on both $m_D$ (and thus $m_{\rm DM}$) and the Yukawa coupling $y$. For $y \lesssim 1$, the contour representing the observed DM contour only slightly deviates from the above value, allowing for DM masses up to $m_{\rm DM}\lesssim 1.2$~TeV. In Sec.~\ref{sec:msbound}, however, we had argued that a strong metastability bound requires \textit{both} a value of $m_S$ close to the EW scale \textit{and} a sizeable Yukawa coupling, $y\sim \mathcal{O}(1)$. As made evident in Fig.~\ref{fig:relic-density}, this regime corresponds to significantly increased DM masses, $m_{\rm DM}\sim 1.2-2$~TeV. Altogether, this implies that observing a DM particle with a mass within this range automatically implies a strong Higgs mass metastability bound in our singlet-doublet scenario.

\section{Electroweak vacuum lifetime and heavy new physics scale}\label{sec:lifetime}

In Sec.~\ref{sec:DMMSB}, we have argued that the model under consideration in this work can plausibly avoid the hierarchy problem and reproduce the observed DM relic density for fermion masses of $\mathcal{O}$(TeV) and $\mathcal{O}(1)$ Yukawa couplings. The former of these observations relies on the effect of the Yukawa couplings on the Higgs quartic coupling, ultimately lowering the instability scale $\mu_I$ due to radiative effects. This may appear to destabilise the EW vacuum to the point where its lifetime becomes shorter than the observed age of the universe. However, our description so far is as an EFT below some heavy scale at which we expect a more UV-complete theory with additional BSM degrees of freedom at the EFT cut-off to (partially) restore stability. In the remainder of this section we will show how its ability to do so can be used to to constrain the scale of this heavy new physics. In the process we will combine and refine the mostly model-independent arguments previously given in Refs.~\cite{Khoury:2021zao,Steingasser:2022yqx,Chauhan:2023pur,Benevedes:2025qwt}.

The lifetime of the EW vacuum is defined as the time at which there is an $\mathcal{O}(1)$-probability that a bubble of true vacuum has nucleated in the past light cone $\mathcal{P}(t)$ of any given observer. In terms of the bubble nucleation rate per unit volume $\Gamma/V$, this amounts to
\begin{equation}
	 \int_{\mathcal{P}} \text{d}^4 x \ \frac{\Gamma}{V}\sim 1.
    \label{eq:total-probability}
\end{equation}
In other words, the ongoing existence of the Universe requires that the left hand side of this expression is significantly smaller than $1$ for our own past light cone. The volume of the latter is given by~\cite{Buttazzo:2013uya}
\begin{equation}
    V_{\mathcal{P}}=\SI{2.2e163}{\GeV^{-4}}.
    \label{eq:VolumeLightcone}
\end{equation}
Now using that the bubble nucleation rate per unit volume is to leading order controlled by the Euclidean action $S_E$ of some suitable instanton with typical scale $\mu_s$, $\Gamma /V \sim \mu_s^4 e^{-S_E}$, this translates to a lower bound on the Euclidean action of this instanton,
\begin{equation}\label{eq:SEbound}
    S_E > 410 + 4 \ln \left( \frac{\mu_S}{\si{TeV}} \right).
\end{equation}
The form of $S_E$ now depends on the relevant degrees of freedom at the scale of interest $\mu_s$, which itself is determined by the form of the potential in this regime.

To ensure that our previous analysis remains independent of the details of the UV completion, we will restrict ourselves to scenarios in which the scale of the stabilising physics $\Lambda$ lies above the instability scale, $\mu_I <\Lambda$. To obtain an upper bound on $\Lambda$, we first consider the limit $\Lambda \gg \mu_s $. In this limit, the Higgs' mass term as well as higher-dimensional contributions arising from the UV completion are negligible,
\begin{align}\label{eq:VIR}
    V= - \frac{|\lambda|}{4} h^4.
\end{align}
The classical scale invariance of this potential implies the existence of an infinite number of decay channels represented through a one-parameter family of instantons~\cite{Andreassen:2017rzq},
\begin{align}
    h_R (\rho)= \sqrt{\frac{8}{|\lambda|}}\frac{1}{1+\rho^2/R^2} \frac{1}{R} \quad \text{with} \quad  S_E[h_R]= \frac{8 \pi^2}{3 |\lambda|}. \label{eq:SE}
\end{align}
Here, $\rho$ is the Euclidean spacetime radius while $R$ labels different members of the one-parameter family. The scale invariance is, however, broken by the RG-scale dependence of $\lambda$. Thus, the tunnelling rate is dominated by the instanton minimizing $S_E$,
\begin{align}
    \frac{\Gamma}{V}\sim \int {\rm d} R \ \exp \left(- \frac{8 \pi^2}{3 |\lambda (\mu=1/R) |}\right) \sim \exp \left(- \frac{8 \pi^2}{3 |\lambda (\mu_*) |}\right),
\end{align}
where the scale $\mu_*$ is obtained by minimizing $S_E$ with respect to $R$. Eq.~\eqref{eq:SE} then implies that $\mu_*$ corresponds to the scale where $\lambda$ reaches its minimum. The bound~\eqref{eq:SEbound} thus translates to a lower bound on $\lambda$ at its minimum~\cite{Isidori:2001bm, Benevedes:2025qwt},
\begin{align}\label{eq:lambdabound}
    \lambda (\mu_*) \gtrsim -0.06 .
\end{align}
In the absence of a UV completion of the Higgs sector, this bound would indeed be violated for the entire parameter space of interest from the point of view of the metastability bound. For notational simplicity, we introduce the scale $\mu_{\rm id}$ by $\lambda (\mu_{\rm id})=-0.06$. 

\begin{figure}[t!]
    \centering
    \includegraphics[width = 0.48\textwidth]{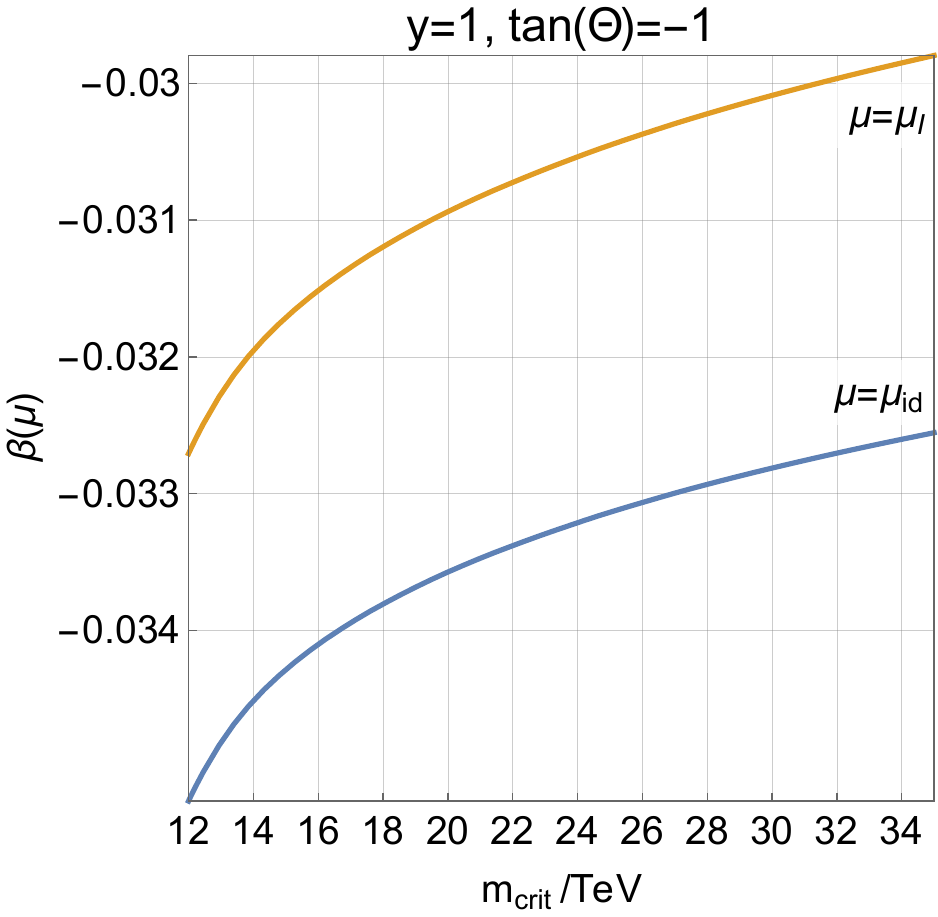}
    \includegraphics[width = 0.48\textwidth]{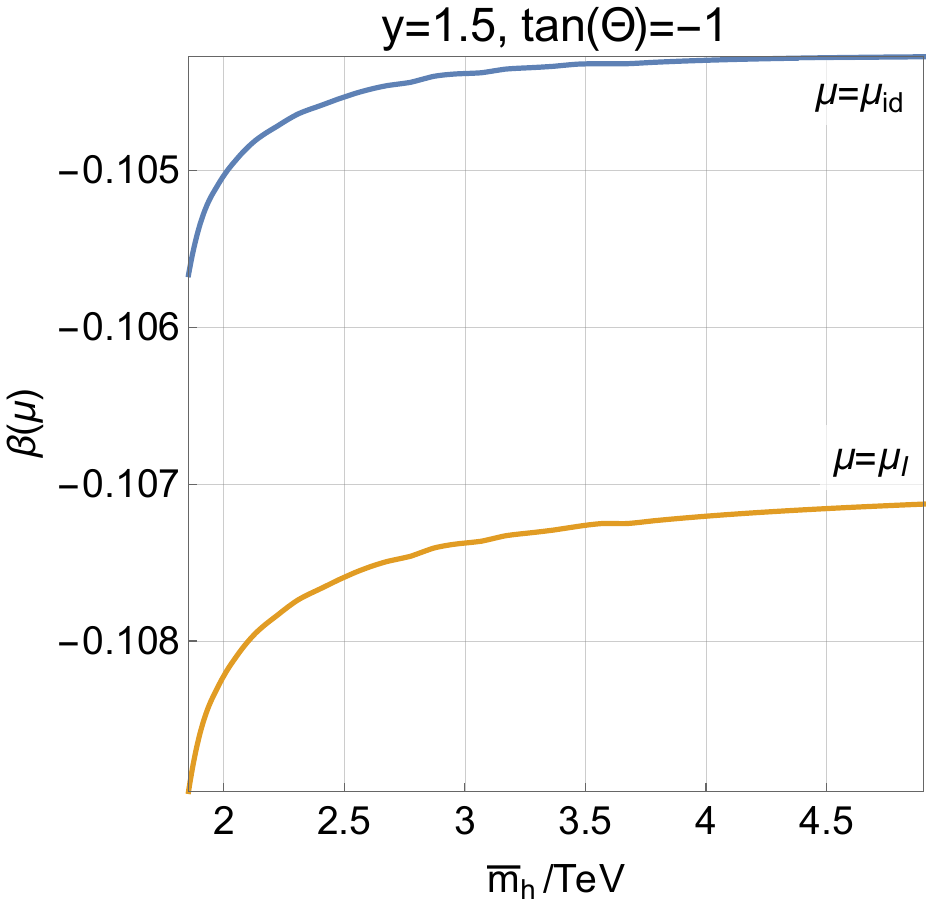}
    \caption{Assuming the complete destabilisation of the vacuum to be avoided through new physics affecting the running implies a lower bound on the contributions of this new physics to the beta function of $\lambda$, depending on its scale where it becomes active. 
    \textbf{Left:} The $\beta$ function at the two reference scales $\mu_I$ and $\mu_{\rm id}$ as function of the metastability bound along the $100\%$ DM contour with $y=1$ and $\tan{\theta} =-1$. \textbf{Right:} The same functions for $y=1.5$.}
    \label{fig:beta}
\end{figure}

\begin{figure}[t!]
    \centering
    \includegraphics[width = 0.48\textwidth]{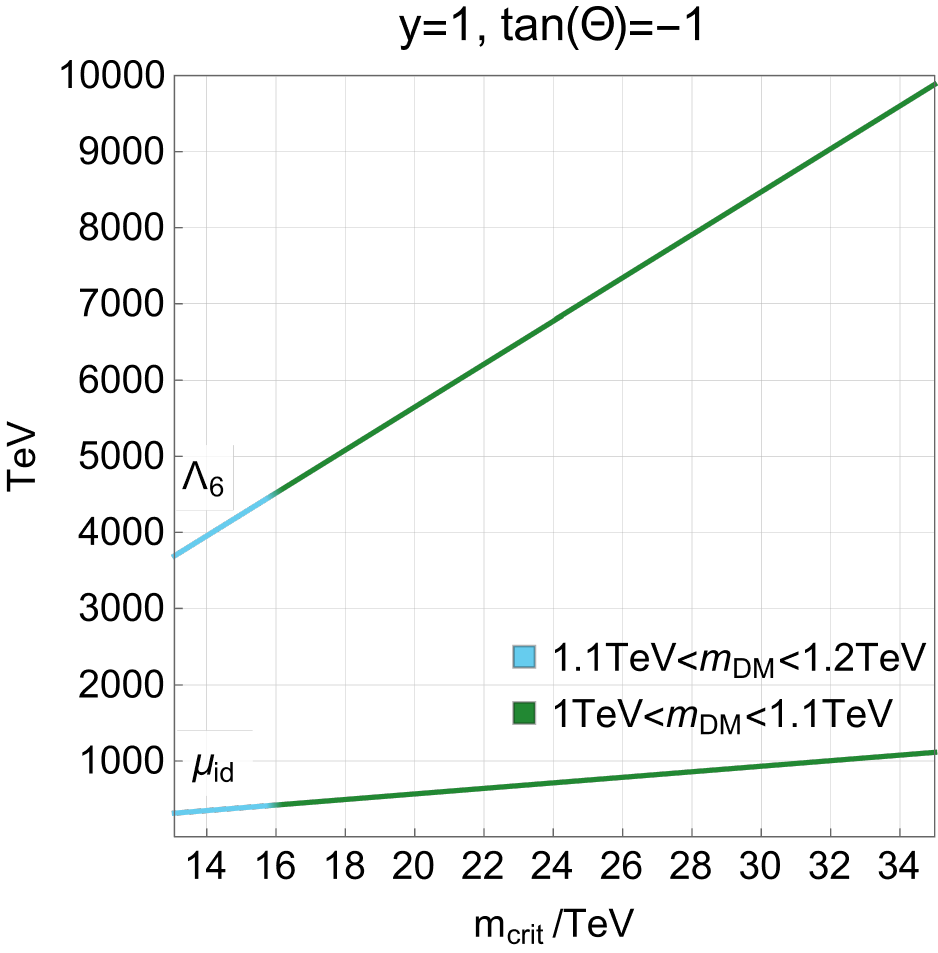}
    \includegraphics[width = 0.48\textwidth]{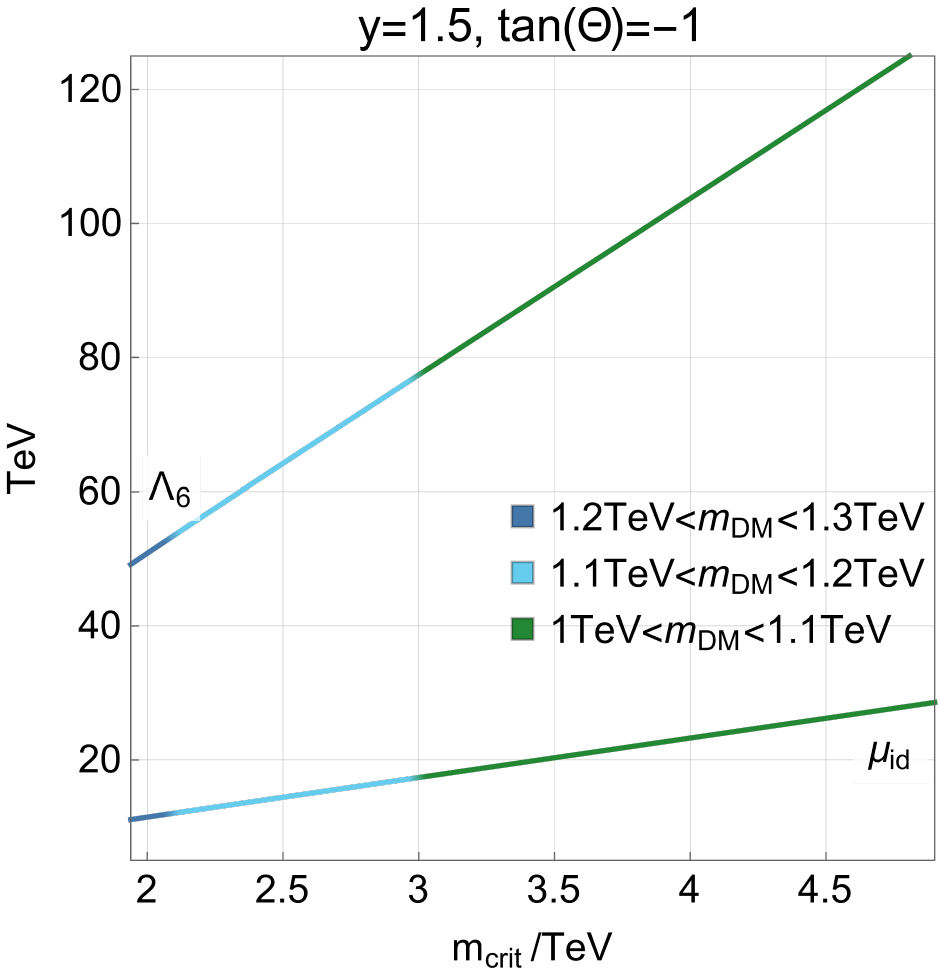}
    \caption{\textbf{Left:} The relevant scales $\mu_{\rm id}$ and $\Lambda_6$ as function of the metastability bound along the $100\%$ DM contour with $y=1$ and $\tan{\theta} =-1$. In the limit of a weak metastability bound, the DM mass converges against $\sim1090$~GeV, whereas DM masses above the threshold of $1.1$ correspond to regimes with a tighter bound.  \textbf{Right:} The same functions for $y=1.5$. A tight bound on the Higgs mass again corresponds to a DM mass above the asymptotic value of $m_{\rm DM}\sim 1085$~GeV. Following our previous arguments, we find that a strong Higgs mass metastability bound also implies a tighter upper bound on the stabilisation scale.}
    \label{fig:scales2}
\end{figure}

Keeping the lifetime of the EW vacuum above the current age of the Universe thus requires either changing the running of the quartic coupling $\lambda$, or changing the shape of the potential, and thus the Euclidean action. The former case is only feasible if both $\Lambda < \mu_{\rm id}$ and the new physics contribution to the beta function of $\lambda$ is sufficiently large and positive to stabilise $\lambda$ or drive it back to positive values. To allow for an estimate of this effect, we show in Fig.~\ref{fig:beta} the values of $\beta_\lambda (\mu_{\rm in})$ and $\beta_\lambda (\mu_{\rm id})$ along the contours corresponding to the observed DM abundance shown in Fig.~\ref{fig:relic-density}. For the value of $\mu_{\rm id}$ along these same contours, see Fig.~\ref{fig:scales2}.

The situation is more involved if the EW vacuum is (partially) stabilised by a deformation of the potential. While this can also trivially be achieved if $\Lambda < \mu_{\rm id}$, the EW vacuum can also be stabilised through the effects of new physics at even higher scales. To quantify this effect, we consider the effective potential along the field-space contour formed by the instanton profile. Parametrising displacements along this contour by an appropriately normalised field $\eta$, this allows us to describe the tunnelling process to leading order  through an effective theory for $\eta$,
\begin{gather} \label{eq:SVeta}
    \mathcal{L}_{\rm inst} (\eta)=\frac{1}{2} \partial_\mu \eta  \partial^\mu \eta  - V_{\rm eff}(\eta), \quad \text{where} \quad
    V_{\rm eff}(\eta)= \frac{1}{4} \lambda_{\rm eff}(\eta) \eta^4+ \Delta V(\eta).
\end{gather}
As argued in Ref.~\cite{Enguita:2025ybx}, in the presence of a significant hierarchy we can identify the field $\eta$ to leading order with the low-energy Higgs field, implying that the effective coupling $\lambda_{\rm eff}$ in Eq.~\eqref{eq:SVeta} also agrees with the Higgs self-coupling. For the purpose of identifying an upper bound on the scale of stabilising physics $\Lambda$, it would appear natural to attempt a perturbative treatment of the new physics terms collected in $\Delta V$. This would imply that, to leading order, the tunnelling process is still dominated by the one-parameter family of bounce solutions in Eq.~\eqref{eq:SE}, with the corresponding Euclidean action,
\begin{gather}
    S_E=\int {\rm d}^4 x\ \mathcal{L}_{\rm inst}(\eta= \phi_R)= \frac{8 \pi^2}{3 |\lambda|}+\Delta S_E,\quad \text{where}\quad 
    \Delta S_E= \sum_{n=1}^\infty \frac{s_{2n+4}}{|\lambda|^{n+2}} \left(\frac{1}{R \Lambda}\right)^{2n}.
\end{gather}
The coefficients $s_{2n+4}$ are related to the Wilson coefficients of $\Delta V$ through
\begin{align}
    s_{2n+4}= 8^{n+2} C_{2n+4}\int {\rm d} \rho \ \rho^3 \left(\frac{1}{1+\rho^2/R^2}\right)^{2n+4}.
\end{align}
Similar to the previous case, the integral over the bounce radius $R$ can be performed through a saddle point approximation, implying that the tunnelling rate is dominated by a single bounce whose radius $R_s\equiv \mu_s^{-1}$ minimises $S_E(R)$. It is straightforward to show that this amounts to the condition 
\begin{align}\label{eq:instscaledef}
    \beta_\lambda (\mu_s)\left[\frac{8 \pi^2}{3} +\sum_{n=1}^\infty  s_{2n+4} \frac{(n+2)}{|\lambda|^{n+1}} \left(\frac{\mu_s}{\Lambda}\right)^{2n}\right]=-\sum_{n=1}^\infty  s_{2n+4} \frac{2n}{|\lambda|^n}\left(\frac{\mu_s}{\Lambda}\right)^{2n} .
\end{align}
The effect of the higher-dimensional operators can best be illustrated by considering a single term in the expansion. Taking for concreteness the correction $\propto \eta^6$, Eq.~\eqref{eq:instscaledef} simplifies to
\begin{align}\label{eq:instscale}
    \beta_\lambda (\mu_s)\left[\frac{8 \pi^2}{3} +  s_{6} \frac{3}{|\lambda|^2} \left(\frac{\mu_s}{\Lambda}\right)^{2}\right]=- s_{6} \frac{2}{|\lambda|}\left(\frac{\mu_s}{\Lambda}\right)^{2}
\end{align}
In the absence of the higher-dimensional term, this equation would imply $\beta_\lambda (\mu_s)=0$, reproducing Eq.~\eqref{eq:SE}. If, however, $C_6>0$ --- or, equivalently, $s_6 >0$ --- the right-hand side of Eq.~\eqref{eq:instscale} is negative, whereas the coefficient multiplying $\beta_\lambda (\mu_s)$ remains positive. This requires $\beta_\lambda (\mu_s)<0$, i.e.\@, $\mu_s<\mu_*$, thereby lowering the instanton scale. In other words, the higher-dimensional operators stabilise the vacuum not only \emph{directly} through their contribution to the Euclidean action, but also \emph{indirectly} by changing the RG scale at which the leading-order contribution is evaluated~\cite{Khoury:2021zao,Chauhan:2023pur}. Importantly, this implies the existence of a regime in which the higher-dimensional terms stabilise the vacuum \emph{without} to leading order influencing the shape of the instanton~\eqref{eq:SE}, upholding the validity of our perturbative analysis. 

We can now obtain an order-of-magnitude estimate for the scale $\Lambda$ necessary for this effect to allow for a significant stabilisation by demanding that $\mu_s \lesssim \mu_{\rm id}$. Crucially, this is a \textit{weaker} requirement than demanding the stabilising effect to be sufficient to keep the vacuum lifetime above the current age of the Universe, which would require a full tunnelling computation. However, in the regime of interest to us such a computation would be sensitive to the details of the UV completion, making it model-dependent~\cite{Khoury:2021zao,Enguita:2025ybx}.

Using Eq.~\eqref{eq:instscale} and neglecting the second, subdominant term in the coefficient of $\beta_\lambda$, a significant stabilising effect requires
\begin{align}
    \Lambda\lesssim  \left(\frac{3 s_{2n+4}}{8 \pi^2} |\beta_\lambda (\mu_{\rm id}) | |\lambda (\mu_{\rm id})| \right)^{-\frac{1}{2n}} \cdot \mu_{\rm id} = \left( 2\cdot 10^{-3}s_{2n+4} |\beta_\lambda (\mu_{\rm id}) |  \right)^{-\frac{1}{2n}} \cdot \mu_{\rm id}
\end{align}
for some $n \in \mathbb{N}$. Assuming a generic scaling of the Wilson coefficients, it can be expected that this expression is dominated by the dimension-six term corresponding to $n=1$. Due to both the small numerical coefficient as well as the loop-level suppression from the beta function, this allows indeed for a higher scale of stabilising physics $\Lambda$ compared to the scenario where the stabilisation unfolds exclusively through a change in the running of $\lambda$. 

Assuming for concreteness that the stabilising potential consists only of a dimension-six term $\Delta V= 1/\Lambda_6^2$ (with a possible Wilson coefficient having been absorbed into $\Lambda_6$), we find
\begin{align}
    \Delta S_{E,6}=\frac{128 \pi^2}{5 |\lambda|^3} \frac{1}{(R \Lambda_6)^2} \quad \Leftrightarrow \quad s_6 = \frac{128 \pi^2}{5}.
\end{align}
Thus, the dimension-six operator having a stabilising effect on the vacuum requires
\begin{align}
    \Lambda_6 \lesssim 1.46\frac{\mu_{\rm id}}{\sqrt{|\beta_\lambda (\mu_I)|}} .
\end{align}
We visualise the relation between this scale and the metastability bound in Fig.~\ref{fig:scales2}, also highlighting its elevation relative to $\mu_{\rm id}$. It is worth recalling that a dimension-six operator satisfying this bound would not be sufficient to achieve metastability, but only have a stabilising effect. Thus, it can in general be expected that achieving metastability requires an even lower scale of new physics. In Refs.~\cite{Khoury:2021zao,Benevedes:2024tdq}, it was shown (assuming different BSM fermions lowering the instability scale) that in regimes leading to a strong metastability bound this effect is large enough for back reactions on the form of the instanton as given in Eq.~\eqref{eq:SE} to become significant. Thus, a reliable lifetime calculation would require a complete numerical treatment, and thus specifying a concrete model~\cite{Benevedes:2024tdq}. Finally, we also note the possibility that some non-generic new physics keeps $\lambda$ sufficiently small and negative, between -0.06 and 0, with $\beta_\lambda$ sufficiently small, over a large energy range such that the UV completion can be postponed to much heavier scales than naively expected.

\section{Experimental and observational constraints}\label{sec:constraints}
We briefly summarise relevant constraints for the SDM and reinterpret searches for electroweakinos. For similar compiled constraints for the SDM in previous works, see e.g.\@ Refs.~\cite{Enberg:2007rp,DEramo:2007anh,Cheung:2013dua,Cohen:2011ec,Bhattiprolu:2025beq,Calibbi:2015nha}. 

\subsection{Direct Detection Constraints}
Direct detection experiments probe DM models by constraining DM scattering off nucleons. We use \textsc{micrOMEGAs~v6.2.4} to compute the cross sections for WIMP-nucleon interactions, and employ the LZ2024 results \cite{LZ:2024zvo} as the most competitive bounds on spin-independent (SI) and spin-dependent (SD) interactions.

\begin{figure}[t]
    \centering
    \includegraphics{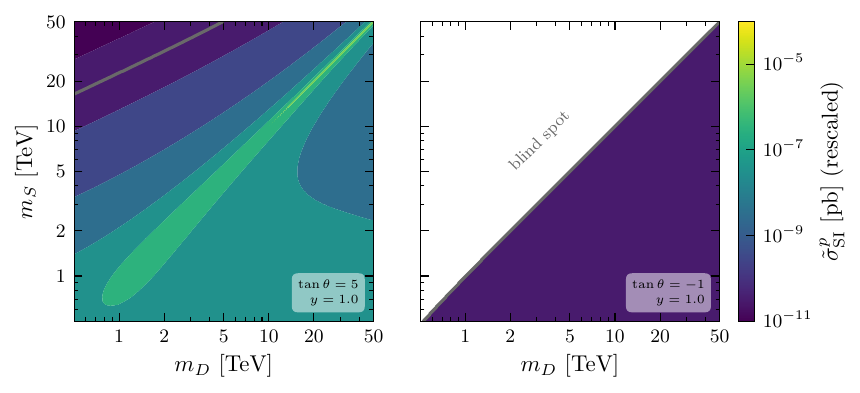}
    \caption{The rescaled spin-independent cross section $\Tilde{\sigma}_{p}$ measured in direct detection experiments for WIMP-proton interactions (comparable to WIMP-neutron constraints) are shown for $y=1.0$ as a function of the doublet and singlet mass. The current experimental limits are indicated by a solid grey line. \textbf{Left:} For $\tan{\theta} = 5$ the, current limits exclude singlet masses below $\SI{20}{TeV}$. \textbf{Right:} In the blind spot with $\tan{\theta}=-1$, the mostly-singlet parameter space below the main diagonal is excluded. The WIMP-nucleon interactions vanish in the mostly-doublet regime in the blind spot (above the main diagonal).}
    \label{fig:direct-detection-SI}
\end{figure}

\paragraph{Spin-independent direct detection constraints}
The LZ2024 results imply an upper limit on the spin-independent cross section for DM-nucleon scattering for DM masses in the $\SI{100}{GeV}$ to $\SI{10}{TeV}$ range of
\begin{equation}
    \sigma_\mathrm{SI} \lesssim \SI{3e-11}{pb} \times \left( \frac{m_\mathrm{DM}}{\SI{1}{TeV}} \right) \times \left( \frac{0.120}{\Omega h^2} \right) .
\end{equation}
As shown in Fig.~\ref{fig:direct-detection-SI}, these constraints exclude the mostly-singlet DM parameter space except for blind spots; see for example Ref.~\cite{Bhattiprolu:2025beq} for a detailed discussion. In the mostly-doublet DM parameter space, the doublet mass is forced to be about $\SI{1.1}{TeV}$ to reproduce the correct DM abundance and is largely independent of the value of the singlet mass parameter for $m_S > m_D$, see also Fig.~\ref{fig:relic-density}. Fig.~\ref{fig:direct-detection-SI} indicates that away from the blind spot regime the direct detection constraints force a large splitting $m_S/m_D \gg 1$.
Here, we show the rescaled cross sections relevant for direct detection experiments,
\begin{align}
    \Tilde{\sigma} = \sigma \times \left( \frac{\SI{1}{TeV}}{m_{\rm DM}} \right) \times \left(\frac{\Omega_{\rm DM}h^2}{0.120}\right) .
\end{align}

\paragraph{Spin-dependent direct detection constraints}
The spin-dependent constraints from LZ2024~\cite{LZ:2024zvo} are generally weaker than the spin-independent constraints for heavy dark matter. The upper bounds on the spin-dependent cross sections for scattering off protons and neutrons, respectively, from LZ2024 are
\begin{align}
    \sigma_\mathrm{SD}^p \lesssim \SI{1.7e-4}{pb} \times \left( \frac{m_\mathrm{DM}}{\SI{1}{TeV}} \right) \times \left( \frac{0.120}{\Omega h^2} \right) ,\\
    \sigma_\mathrm{SD}^n \lesssim \SI{5.3e-6}{pb} \times \left( \frac{m_\mathrm{DM}}{\SI{1}{TeV}} \right) \times \left( \frac{0.120}{\Omega h^2} \right)  ,
\end{align}
which are valid in the mass range $\SI{100}{GeV}$ to $\SI{100}{TeV}$. The spin-dependent limits from proton scattering do not constrain the parameter space above $\SI{500}{GeV}$ considered throughout this work. The spin-dependent limits for scattering off neutrons generally exclude the parameter space of lower and aligned doublet and singlet masses. However, the SI limits impose much stronger constraints on the parameter space. We therefore only show the spin-independent constraints representative for the direct detection limits in Fig.~\ref{fig:direct-detection-SI}.

\subsection{Indirect Detection Constraints}
Indirect detection DM experiments constrain the annihilation cross section in the late Universe, in particular by measuring the photon flux from the Galactic centre or dwarf galaxies induced by DM annihilation into SM final states. 
DM annihilations in the SDM can generally occur into all EW states at tree-level, though diboson final states are the dominant annihilation channels in the late Universe, in particular $W^{+}W^{-}$, $ZZ$ and also $hh$.
The Yukawa interactions play a subdominant role as Higgs-mediated annihilation into SM fermion is generally $p$-wave for Majorana DM, and correspondingly velocity-suppressed in the late Universe. Therefore, the annihilation cross section is dominated by gauge interactions.
Indirect detection searches for the fermionic annihilation channels of electroweakinos, in particular higgsino DM, can then be recast for the SDM, having similar gauge interactions. Similarly, limits on higgsino DM from diboson annihilation channels can be recast for the SDM. Importantly, the mixing between the Majorana gauge eigenstates controls the coupling constants for the DM gauge interactions. However, note that even in the blind spot, while the diagonal couplings of Higgs and Z bosons to the DM candidate vanish, t-channel exchange of the heavier Majorana or Dirac states with non-vanishing couplings still leads to a sizeable annihilation cross section, unlike for direct detection constraints.

Current constraints from indirect searches exclude doublet masses up to $\sim \SI{300}{GeV}$ in the mostly-doublet regime, largely independent of the singlet mass. See also e.g.\@ Ref.~\cite{Calibbi:2015nha}. Current limits therefore do not probe the parameter space above the EW scale considered here.
Upcoming indirect detection experiments may be able to further probe the parameter space. As argued recently in Ref.~\cite{Abe:2025lci}, CTAO-North may be able to discover or exclude higgsino DM. For the SDM, this result implies that indirect detection searches would probe the parameter space up to doublet masses of at least up to $\SI{1.1}{TeV}$.

\subsection{Collider Searches}
The singlet-doublet model is also constrained by collider experiments through indirect and direct probes. Electroweak Precision Observables (EWPO) constrain the singlet-doublet model indirectly, and dedicated direct searches for the singlet-doublet model and electroweakinos have been conducted.

\paragraph{Electroweak Precision Tests}
EWPO are also sensitive to the fermion extension, and can be combined into the Peskin-Takeuchi parameters measuring the amount of induced custodial isospin violation via the oblique $T$ parameter, and chiral structure of the new sector through the oblique $S$ parameter. In the vector-like extension at hand, the $T$ parameter therefore gives the leading sensitivity to the BSM effects.
The $S$ parameter is proportional to the Yukawa couplings squared but suppressed by $v^2/m_1^2$ and therefore quickly vanishes at large masses.
$T$ exactly vanishes in the alignment limit $y_1 = y_2$ due to the custodial symmetry restoration.
See also Refs.~\cite{Calibbi:2015nha,DEramo:2007anh,Barbieri:2006bg} for further details.

\begin{figure}[t]
    \centering
    \includegraphics{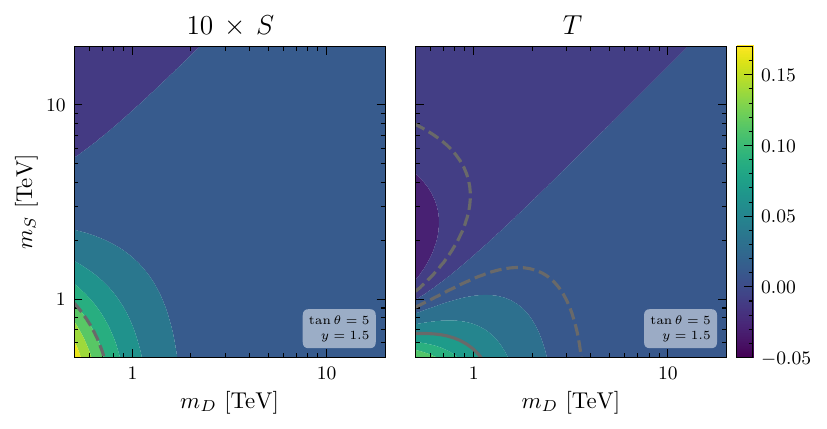}
    \caption{Electroweak precision observables $S$ (left) and $T$ (right) for $y\!=\!1.5$ and $\tan{\theta} \! = \! 5$. \textbf{Left:} The oblique $S$ parameter rescaled by a factor of 10 is shown.  Projected limits from future colliders are represented by grey dashed lines. Current constraints are not visible in this parameter space. \textbf{Right:} Current constraints (solid grey) on the oblique $T$ parameter exclude masses up to $\sim \SI{600}{GeV}$. Projected limits (dashed grey) can probe singlet and doublet masses below a few TeV away from the degenerate limit, $m_S \not\approx m_D$.}
    \label{fig:EWPO}
\end{figure}

We show the oblique parameters $S$ and $T$ for benchmark values of $y=1.5$ and $\tan{\theta}=5$ in Fig.~\ref{fig:EWPO} in the singlet-doublet mass plane. Current bounds on the $S$ parameter are irrelevant for the parameter space shown. Present limits on the $T$ parameter (assuming $U=0$) of $-0.06 < T < 0.06$ \cite{ParticleDataGroup:2024cfk} constrain the SDM for singlet and doublet masses only up to $\sim \SI{600}{GeV}$.
As an optimistic benchmark, we take the sensitivity of future colliders to the $S$ and $T$ oblique parameters of $1\%$ \cite{Fan:2014vta,deBlas:2013gla,deBlas:2016nqo,deBlas:2025gyz}. Projecting constraints on the parameter space assuming a best fit value of $S=T=0$ at future colliders are shown in dashed grey in Fig.~\ref{fig:EWPO}. Bounds on the $S$ parameter will only constrain the parameter space for masses below $\SI{1}{TeV}$, while projections for the $T$ parameter will cover singlet and doublet masses of up to a few TeV, but quickly lose sensitivity if $m_S \sim m_D$. Restricting the masses to reproduce the observed relic abundance for the benchmark Yukawa couplings therefore only gives mild constraints on the model, even in the most optimistic case for future colliders, and is insensitive to the scenario of $m_S \gg m_D \approx \SI{1.1}{TeV}$.

\paragraph{Direct searches}
The SDM can also be probed directly, generally via production of any of the SDM states, at colliders through missing energy searches, disappearing tracks or detection of the charged Dirac state.
While no dedicated direct searches are available for the SDM, in particular results from electroweakino searches can be recast. Presently, several designated direct searches for electroweakinos have been conducted for LHC runs and the FCC-hh, for example Refs.~\cite{ATLAS:2019lff,ATLAS:2022hbt,ATLAS:2021yqv,ATLAS:2022zwa,CMS:2024gyw,CMS:2022sfi,Golling:2016gvc,Gori:2014oua,Calibbi:2014lga}. These studies analysed the production of electroweakino pairs with subsequent decays into observable multi-lepton final states. Studies focussing on higgsino DM analogous to our mostly-doublet DM candidate can be used as conservative estimates given the larger Yukawa coupling in the SDM compared to the supersymmetric case, though the precise constraints depend on the mixing among the Majorana fermions. Previous studies of the SDM, notably Ref.~\cite{Calibbi:2015nha,Bhattiprolu:2025beq}, have recast limits from electroweakino searches, indicating that these constraints are irrelevant for the parameter space considered in this work.

While existing searches do not pose constraints on the motivated part of the SDM parameter space, future colliders will be able to comprehensively probe the SDM. Limits on the neutral fermions will generally be dependent on the mixing between the gauge eigenstates, and are weakened in blind spots of the parameter space. However, the charged Dirac fermion has universal EW couplings over the parameter space and couples to all EW gauge bosons. To feature the observed relic abundance, the electrically charged Dirac fermion must have a mass of $\SIrange[range-phrase = {-}, range-units = single]{1}{1.5}{TeV}$, which can be searched for at high-energy colliders. Such a state would for example be detectable at a $\SI{3}{TeV}$ muon collider~\cite{AlAli:2021let}, and FCC-hh~\cite{Arkani-Hamed:2015vfh,FCC:2018byv,FCC:2025lpp}. Therefore, future colliders can readily assess whether the SDM can account for dark matter. We leave for future work a study of the interesting complementarity with indirect searches for probing the SDM. 
\begin{figure}[t]
    \centering
    \includegraphics[scale=1]{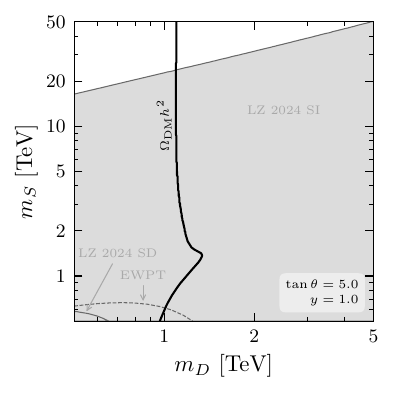}
    \includegraphics[scale=1]{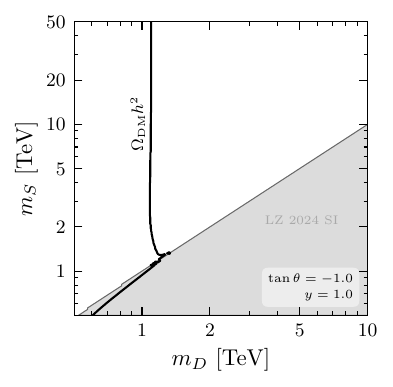}
    \caption{Compiled constraints on the SDM parameter space for $y=1$ from direct detection and electroweak precision measurements. The relic abundance contour corresponding to the observed abundance is marked in black, with the region to the right of it excluded by overproduction of DM for $y=1$. \textbf{Left:} The SDM with $\tan{\theta}=5$ is excluded for a singlet mass below at least $\sim \SI{20}{TeV}$ through SI direct detection experiments. SD direct detection limits and projections for EWPT are subdominant.
    \textbf{Right:} In the blind-spot with $\tan{\theta}=-1$, SI direct detection constraints exclude the entire mostly-singlet regime of the SDM.}
    \label{fig:compiled-constraints}
\end{figure}

\section{Conclusion}
\label{sec:conclusion}

Naturalness may not apply in the expected way to the Higgs. Yet, as a guiding principle for BSM physics, it remains a valuable framework for understanding what nature may be telling us. After all, the principle of relativity did not cease to be relevant following null results in the search for the aether associated with it; it simply did not apply in the way expected at the time. Just as the Michelson-Morley experiment led to a more refined conception of relativity, so may the absence of new physics at the LHC point towards a deeper understanding of the hierarchy problem. 

If the Higgs boson is the only fundamental scalar that is tuned to be light for some as yet unknown reason, then naturalness predicts that all other elementary bosons (at least the fundamental scalars and gauge bosons that get their masses from scalars) are heavy. VL fermions are then amongst the few remaining possibilities for accessibly light new physics~\footnote{Axions are another possibility for naturally light new physics. The interplay between an axion and Higgs criticality has been studied in Ref.~\cite{Detering:2024vxs}.}. VL fermions with Yukawa couplings to the Higgs contribute negatively to the running of the quartic of the Higgs potential. Bosons contribute positively, but are expected to be decoupled to the natural cut-off at heavier scales. A split spectrum of BSM particles could then naturally lead to a metastable Higgs potential with an instability scale not far above the EW scale. 

The central observation of our work is that a natural split spectrum with a low vacuum instability scale coincides with both the thermal WIMP DM paradigm and a cosmological solution to the hierarchy problem motivated by SOC. The WIMP miracle provides independent motivation for VL fermions at the EW scale; naturalness, in the form of a split spectrum, motivates the expectation of a TeV-scale vacuum instability; finally, cosmological vacuum selection relates the Higgs mass to this boundary. These three a priori unrelated ideas come together to form a coherent picture for BSM physics at the TeV scale and provide an alternative explanation for the origin of the EW scale. The role of WIMP DM here is central to setting the scale of the Higgs mass. 

It may be argued that our WIMP DM connection to the hierarchy problem is no different to those found in symmetry-based solutions protecting the Higgs mass, since one could make the similar observation for supersymmetry or composite Higgs models that they contain WIMP DM candidates and the EW scale then follows by virtue of the WIMP miracle. However, in symmetry-based explanations of the EW scale, the mass of the Higgs and the WIMP DM embedded in those models are merely indirectly tied together by both being at the scale at which the symmetry protection mechanism is in effect. In our case, the Higgs mass is originally decoupled and naturally heavy but becomes dynamically driven in the early universe to be close to that of the WIMP DM.

We illustrated our mechanism with a simplified model of singlet-doublet WIMP DM. To evade stringent direct detection limits, we focused on the $m_S > m_D$ parameter space in the blind spot region of suppressed DM couplings to the $Z$ and Higgs boson. The thermal DM mass in this case is at least $\sim 1.1$ TeV. The relic density is largely independent of the value of the Yukawa coupling since it is fixed by annihilation through the weak gauge boson coupling. However, we find that a low enough Higgs metastability bound requires the DM Yukawa coupling to be at least $y \gtrsim 0.8$. A DM mass above 1.2 TeV corresponds to a Yukawa coupling larger than 1, which would indicate a strong metastability bound on the Higgs mass. This also implies an upper bound on the scale at which heavy new physics must enter to restore vacuum stability (though we note that this scale can be made heavier if a UV justification is provided for keeping the $\beta$ function small over an extended energy range). For a Yukawa coupling in the range $1.5 \gtrsim y \gtrsim 0.8$, we estimated the heavy scale to vary between $\mathcal{O}(10)$ TeV to $\mathcal{O}(10^4)$ TeV for our Universe to be sufficiently long-lived. At this scale, additional bosons are expected to contribute to modifications of the Higgs potential. Additional symmetries will also enter to protect the Higgs mass from quantum corrections above this natural heavy cut-off scale. 

Supersymmetry, for example, may be a natural UV completion of our general scenario at the heavy cut-off. Indeed, the picture of a naturally split spectrum was originally put forward in the context of split supersymmetry models where the requirement for supersymmetry to protect the Higgs mass at the EW scale is abandoned~\cite{Arkani-Hamed:2004ymt, Giudice:2004tc, Arkani-Hamed:2004zhs}. In this framework, the heavy scalar superpartners are constrained by the value of the measured Higgs mass to lie in just the right energy range required by our estimate of the scale of vacuum stability restoration~\cite{Giudice:2011cg, Arkani-Hamed:2012fhg, Bagnaschi:2014rsa, Co:2022jsn}. This could in turn give an independent reason to be in the blind spot region of parameter space~\cite{Cheung:2012qy}. The simplified model we have considered here, however, cannot be straightforwardly embedded in a minimal split supersymmetry extension of the SM. The higgsino Yukawa is fixed by supersymmetry to be $\sim 0.1$, too small for our purpose.
An alternative WIMP DM candidate or non-minimal split supersymmetry model is therefore required, which may also provide a new perspective on the $\mu$ problem of supersymmetry. We leave this interesting model-building exercise for future work.

Regardless of how Higgs criticality and the metastability bound are realised in specific models, an inevitable prediction of the general framework is a modification of the Higgs potential not far above the EW scale. This provides a new, motivated scenario for BSM physics at the TeV scale to target at current and future facilities. Naturally light new physics, for example in the form of VL fermions or axions, are expected within a finite parameter space. Their masses and couplings cannot be arbitrarily decoupled as they must interact sufficiently strongly with the Higgs to induce a vacuum instability at the TeV scale. Future colliders can therefore probe virtually the entire natural parameter space of this scenario. If a VL fermion in the form of WIMP DM is responsible, as suggested in this work, we may further see complementary signals or constraints in DM direct detection experiments and astrophysical observations. Next-generation facilities could then definitively settle the question of whether near-criticality of the EW vacuum and the metastability bound have something or not to do with the hierarchy problem.

\section*{Acknowledgements}
The authors acknowledge Víctor Enguita-Vileta for valuable contributions and collaboration during the early stages of this work. This work was supported by the STFC under UKRI grant ST/X000753/1. Portions of this work were conducted in MIT's \textit{Center for Theoretical Physics - a Leinweber institute} and partially supported by the U.S.~Department of Energy under Contract No.~DE-SC0012567. This project was also supported in part by the Black Hole Initiative at Harvard University, with support from the Gordon and Betty Moore Foundation and the John Templeton Foundation. MD is supported by a Faculty Studentship at King's College London. 
TS also acknowledges partial financial support from the Spanish Research Agency (Agencia Estatal de Investigaci\'on) through the grant IFT Centro de Excelencia Severo Ochoa No CEX2020-001007-S and PID2022-137127NB-I00 funded by 
MCIN/AEI/\allowbreak10.13039/\allowbreak501100011033/\allowbreak FEDER, UE. This project has received funding/support from the European Union's Horizon 2020 research and innovation programme under the Marie Sklodowska-Curie Staff Exchange  grant agreement No.~101086085--ASYMMETRY. The opinions expressed in this publication are those of the authors and do not necessarily reflect the views of these Foundations.

\bibliographystyle{JHEP}
\bibliography{ref}

\end{document}